\documentclass[11pt]{article}
\usepackage{amssymb,amsmath,graphicx,float,wrapfig,fancyhdr,lastpage,url,subcaption,textcomp,caption, authblk}
\usepackage[margin=1in]{geometry}
\begin{document}
\title{\textbf{Characterizing information importance and the effect on the spread in various graph topologies}}
\author[1]{James Flamino\thanks{flamij@rpi.edu}}
\author[2]{Alexander Norman\thanks{normaa@rpi.edu}}
\author[2]{Madison Wyatt\thanks{wyattm@rpi.edu}}
\affil[1]{Department of Physics, Applied Physics, and Astrophysics, Rensselaer Polytechnic Institute, 110 8th St, Troy, NY 12180, USA}
\affil[2]{Department of Mathematical Sciences, Rensselaer Polytechnic Institute, 110 8th St, Troy, NY 12180, USA}
\date{April 25, 2016}
\maketitle
\section*{\hfil Abstract\hfil}
\indent

\indent In this paper we present a thorough analysis of the nature of news in different mediums across the ages, introducing a unique mathematical model to fit the characteristics of information spread. This model enhances the information diffusion model to account for conflicting information and the topical distribution of news in terms of popularity for a given era. We translate this information to a separate graphical node model to determine the spread of a news item given a certain category and relevance factor. The two models are used as a base for a simulation of information dissemination for varying graph topoligies. The simulation is stress-tested and compared against real-world data to prove its relevancy. We are then able to use these simulations to deduce some conclusive statements about the optimization of information spread. \textit{Original paper written for COMAP's 2016 ICM competition.}
\\

\newpage
\section{Introduction}

\subsection{Context and Motivation}
\indent
\indent Over time, the methods we use to communicate have evolved and grown from small networks of groups with limiting connections to the massive array of transmission associated with the internet. In any sense, one thing has remained constant--the inherent practice of seeking and sharing news. Communication Theory suggests this is for a few fundamental reasons: to persuade, to give or provide information, to seek information, and to express emotions. These basic desires drive the ways in which we build our networks and determine the importance of each interaction.\\

\indent We seek to develop a mathematical model of social networks to analyze communication at its core. To do this, we need to understand the modes by which people communicate, the information people convey and accept, the science of implanting news and the effects these all have on the resulting spread of the news. \\

\subsection{Assumptions}
\indent
\indent A few assumptions are taken in order to allow the model to run smoothly.
\begin{itemize}
\item We assume topical distribution trends are uniform throughout a given era. 
\item We assume noteworthy news has overcome the threshold of noise and disregard static traffic of information.
\item We assume a discretized time variable, updating the status of an information string in each node at regular intervals. 
\item We assume that the probability of someone registering and paying attention to some string of data, or indeed sharing this piece of data, along mediums of communication is also random.
\item Finally, we assume that when a node decides to share data, it shares it with all of its neighbors. \\
\end{itemize}
\indent Additional small assumptions specific to pieces of the model will be outlined at appropriate times.  \\

\section{Data Analysis}

\subsection{Trends of News Sources}
\indent	
\indent The reach of information has exploded over the past 150 years. In today's society, information from nearly anywhere can propogate into the mainstream media. These connections and spreads can bring about amazing consequences and this phenomenon is seen across time. By compiling data from census reports, statistical assessments and available databases, we determine the trends of four information transfer mediums: newspaper, radio, television, and the internet. The industrial revolution brought a massive increase in newspaper production so that the number of newspapers and subsequent circulation increased dramatically. In the 20th century, radio and TV began to replace newspapers as major sources of information, and in more recent history, the internet has massively expanded the spread of information, working to replace outdated media outlets as a primary source. Newspaper circulation skyrocketed from 1880 to 1940, until radio broadcasting began to gain influence. In 1920, only about 3 million homes had a radio set, but that number reached nearly 30 million by the end of the 1930s. By 1945, 80\% of homes had a radio set and that number continued to grow into today, in which radios are in virtually every car and every home. The introduction of TV saw an even greater increase with only 0.4\% of households having a set in 1948, to 55.7\% four years later. The trend only continued from there. Finally, the age of the internet has exceeded the trends of the previous mediums and seen unparalleled rapid expansion. These trends can be seen in Figure 1(a-c) depicting the true data found.\\

\begin{figure}
 \centering
  \subcaptionbox{\label{fig1:a}}{\includegraphics[width=2in]{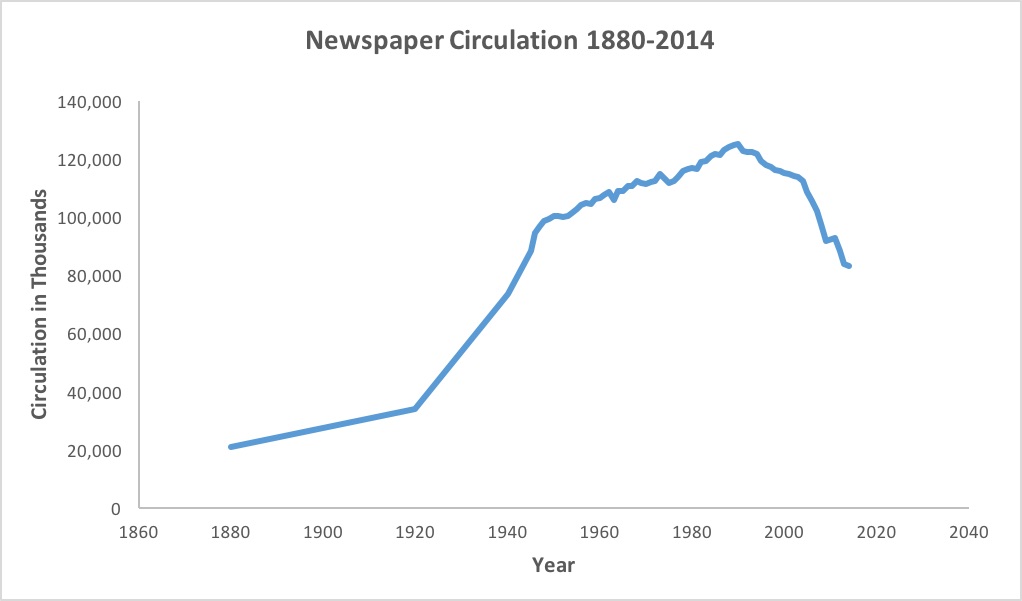}}\hspace{1em}%
  \subcaptionbox{\label{fig1:b}}{\includegraphics[width=1.9in]{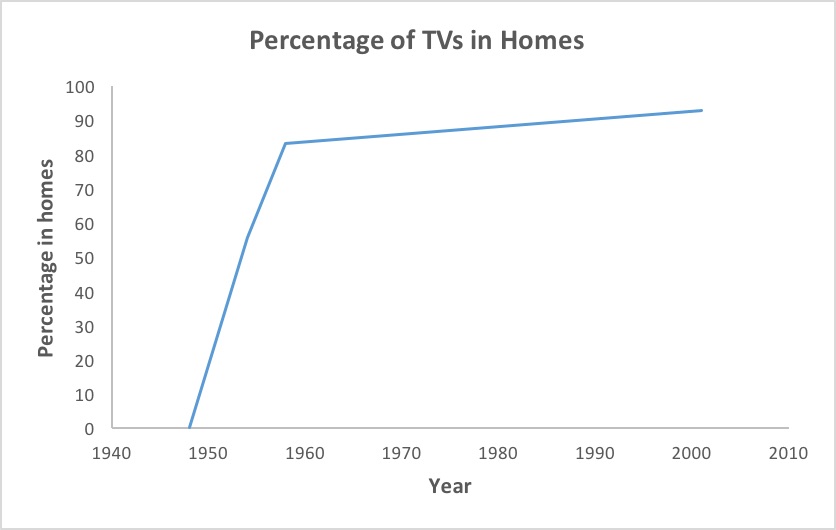}}\hspace{1em}%
  \subcaptionbox{\label{fig1:c}}{\includegraphics[width=2.1in]{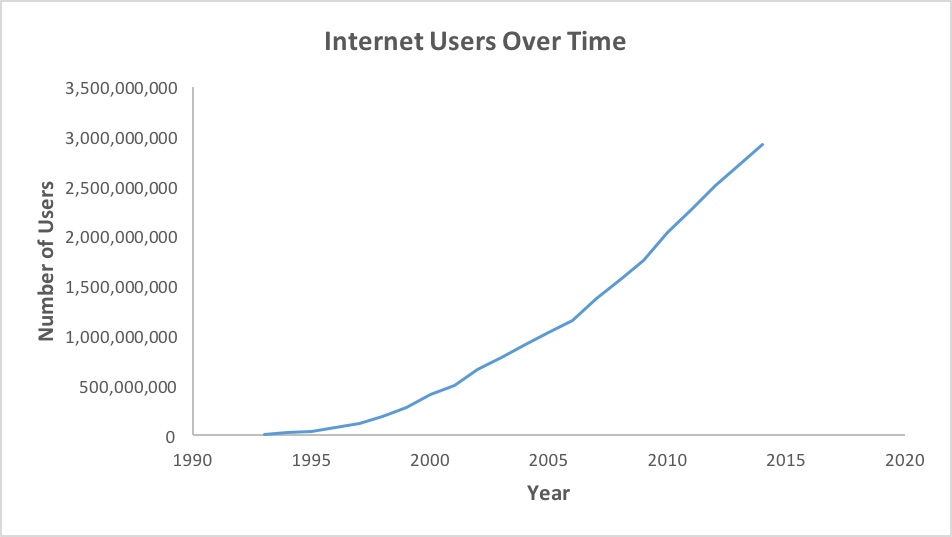}}\hspace{1em}%
  \caption{\small a) Newspaper circulation trends over time. b) Percentage of TVs in US households over time. c) Number of internet users over time. Currently, 40.6\% of the world is connected to the internet.}
\end{figure}

\indent The introduction of each new medium contributed to an eventual downfall of the others. Although there may not be an extinction of any one as they serve to compliment each other, it is shown that the increased usage of a novel medium brings a decrease in relevance of the current and past types. However, innovative sources like the internet combine the functions of newspapers, radio, and television with eNews, internet podcasts and instant streaming, respectively. This could serve to make outdated methods irrelevant and, consequently, obsolete. \\

\subsection{What is news?} 
\indent 
\indent Ultimately, the question at hand is, what makes a news story? How does one piece of information gain relevance over another? For the purpose of this paper, we define ``news" as a piece of information that overcomes two thresholds: the threshold of penetration and the threshold of retention. We are concerned with the information that has the potential to have a wide spread and/or lasting impression. The threshold of penetration is the minimum number of mentions across various sources for a piece of information to catch wind. The threshold of retention is the minimum amount of time a piece of information is relevant enough to be shared. Essentially, in everyday news there is a level of noise of information that the story must overcome to be considered. \\

\indent In order to enforce what qualifies as news, we look at major news stories throughout the past 150 years to get an accurate idea. To compare with actual data, we tracked the volume change of phrases mentioned over periods of time in a database of over 10 million archived newspapers from 1836-1922, in the Washington Post from 1960-1963, and in data compiled from internet use of today. To begin, a phrase was chosen in terms of its relevance to the time period. The volume of mentions of the phrase per day was tracked over a time interval and recorded. \\

\indent We introduce two similar pieces of news and explore their ability to break above the thresholds. First, we examine the volume over time of mentions of the word ``Lincoln" in newspapers from 1864-1866. As seen, there is a consistent level of noise until April 15, 1865, corresponding to the assassination of President Lincoln. 

\begin{figure}[H]
\centering
\includegraphics[scale=.25]{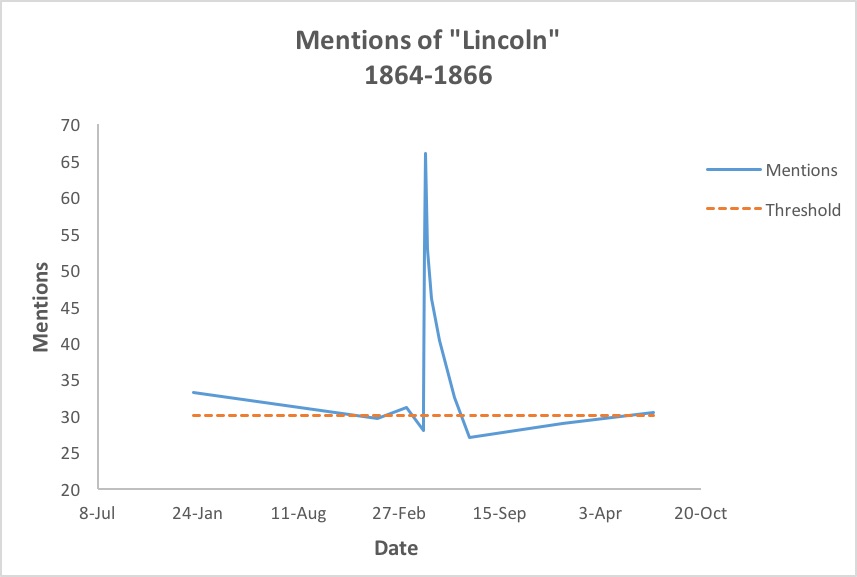} 
\caption{\small Mentions of ``Lincoln" over time in US newspapers from 1864-1866.} 
\end{figure}
\indent After a time period, the volume of mentions returns to the threshold, even delving below as the phrase becomes less relevant over time. This is also seen in a current, but similar scenario from 2012. There is an equal distribution of mentions of ``Osama Bin Laden" and ``Moammar Gaddafi" until a single source released mention of Osama Bin Laden's death into a communication network, signaling a spike in his mentions. This is clearly a piece of information that reflects our definition of news.\\

\begin{figure}[H]
\centering
\includegraphics[scale=.4]{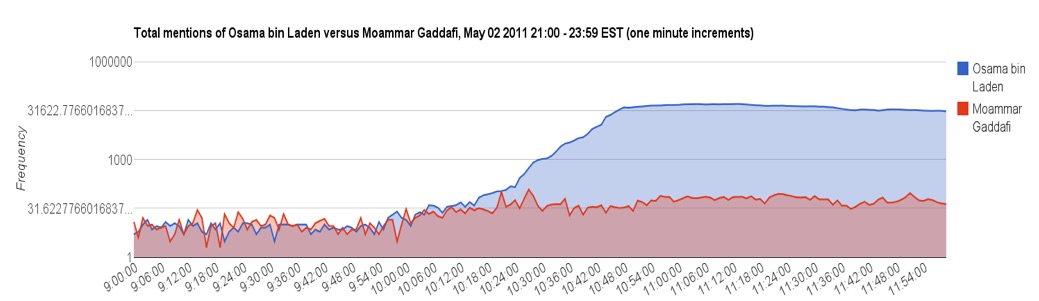} 
\caption{\small Mentions of the phrase ``Osama Bin Laden" over time. The plot of ``Moammar Gaddafi" is included as an equivalence reference.} 
\end{figure}

\indent Across the board, this shape and trend is seen for differing types of news in all eras, topics, and importance to the period. \\

\begin{figure}
 \centering
  \subcaptionbox{\label{fig4:a}}{\includegraphics[width=2in]{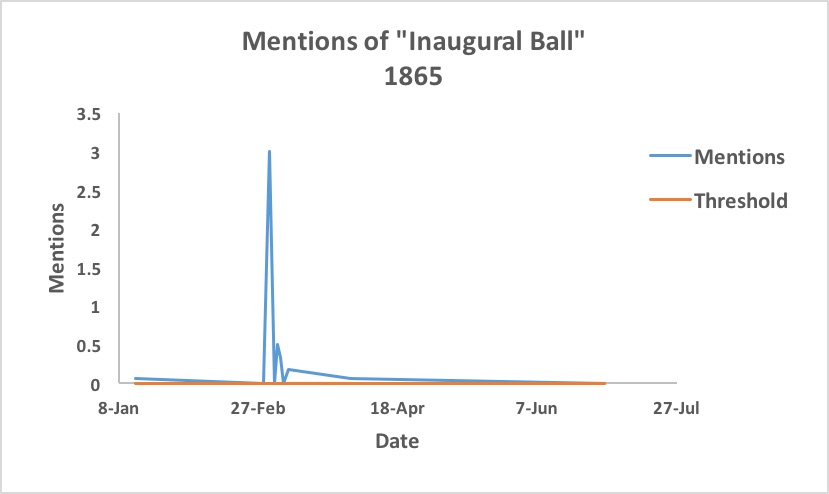}}\hspace{1em}%
  \subcaptionbox{\label{fig4:b}}{\includegraphics[width=1.8in]{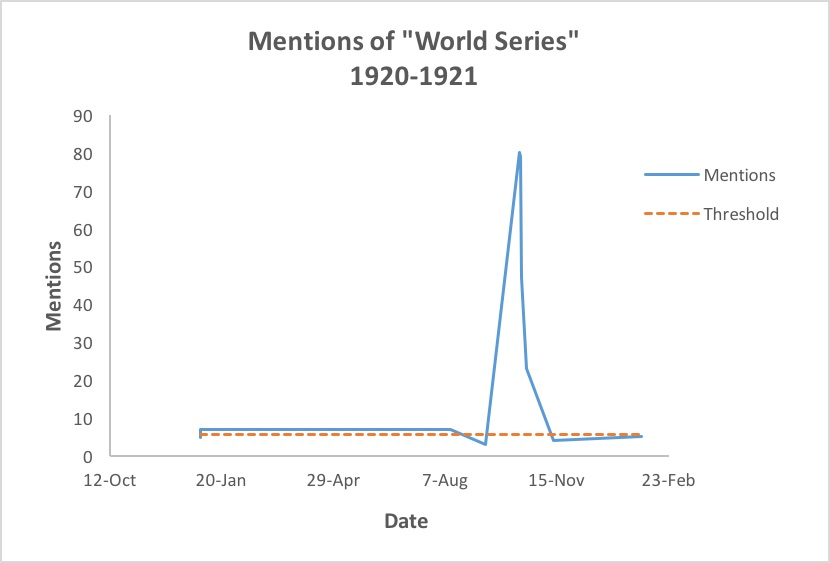}}\hspace{1em}%
  \subcaptionbox{\label{fig4:c}}{\includegraphics[width=1.8in]{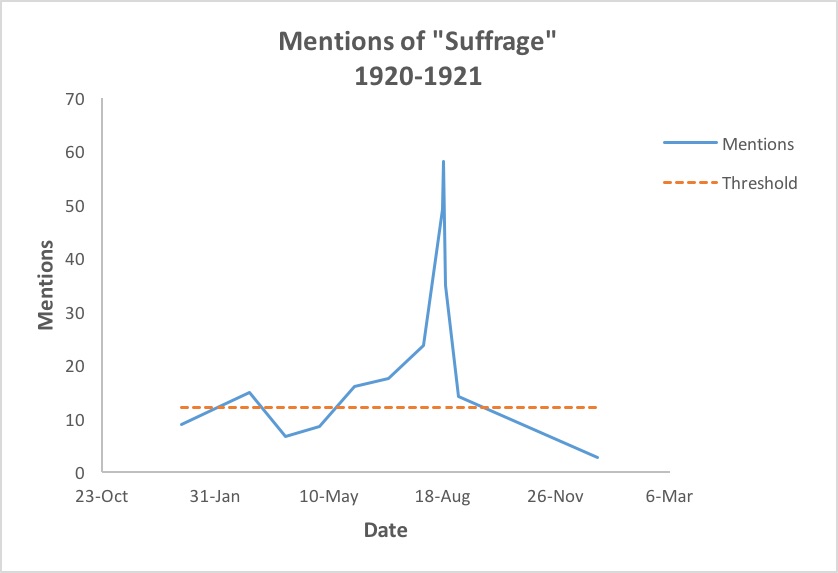}}\hspace{1em}%
  \subcaptionbox{\label{fig4:d}}{\includegraphics[width=1.8in]{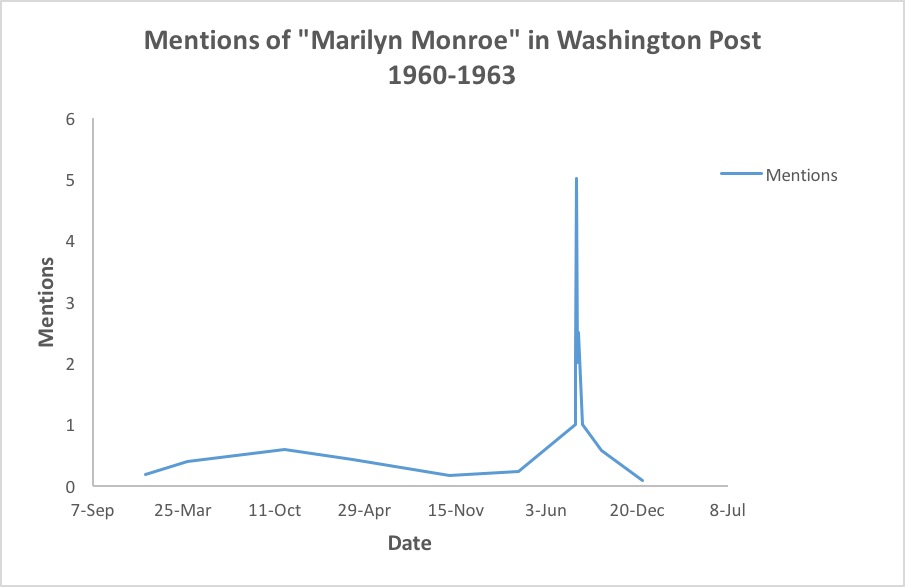}}\hspace{1em}%
  \subcaptionbox{\label{fig4:e}}{\includegraphics[width=1.75in]{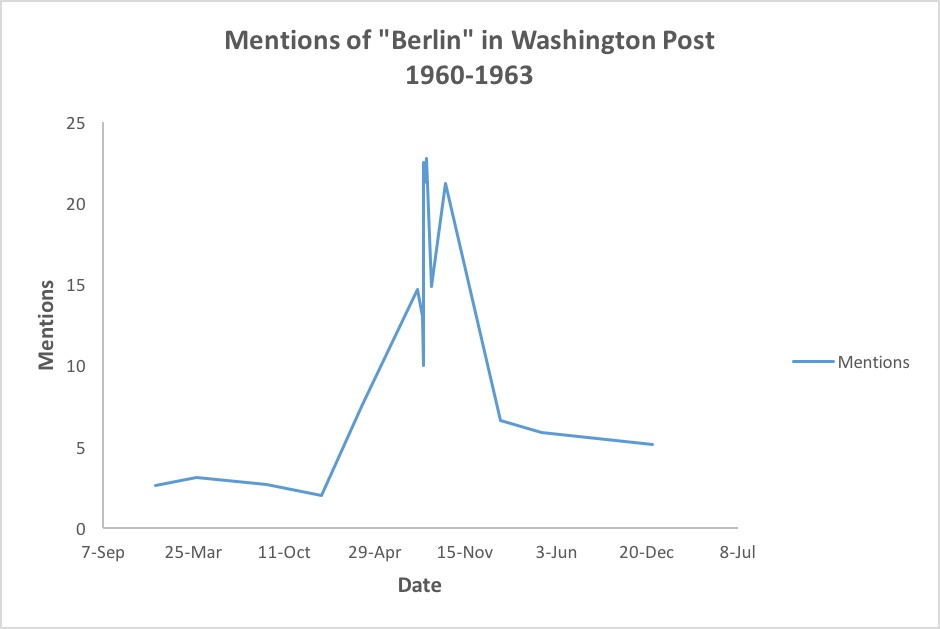}}\hspace{1em}%
  \caption{\small a) Mentions of ``Inaugural Ball" in newspapers from 1864-1866. Peak corresponds to Lincoln\textquotesingle s 2nd Inaugural Ball. b) Mentions of ``World Series" in the early 1920s. Peak corresponds to a Cleveland victory. c) Mentions of ``Suffrage" in the early 1920s. Peak corresponds to the ratification of the 19th Amendment. Notice the dip below threshold after the release as the phrase held less relevance. d) Mentions of ``Marilyn Monroe" from 1960-1962. The peak corresponds to her deadly overdose in 1962. e) Mentions of ``Berlin" from 1960-1962. The peak shows the date of erection of the Berlin Wall. Notice the retention of the news.}
\end{figure}

\subsection{Trends of Topical Distribution}
\indent
\indent Finally, we examine the change over time of topical distribution in news sources. In spreading news, we categorize the highest ranking stories as falling into categories of consumption, politics, and entertainment. That is, the majority of news stories are most relevant in the realm of the aforementioned categories. Again compiling data from census reports on the circulation of different types of newspapers over the years, statistical data on radio and TV broadcast breakdown, and finally examining the top ten news sources with the highest internet traffic for information type, we arrive at a topical distribution for the type of information being spread over time. There is a clear trend away from political news towards entertainment which is reinforced by the comparison of phrase volume distributions within eras. \\
\begin{figure}[H]
 \centering
  \subcaptionbox{Circulation proportion of newspaper type in 1880.\label{fig5:a}}{\includegraphics[width=1.7in]{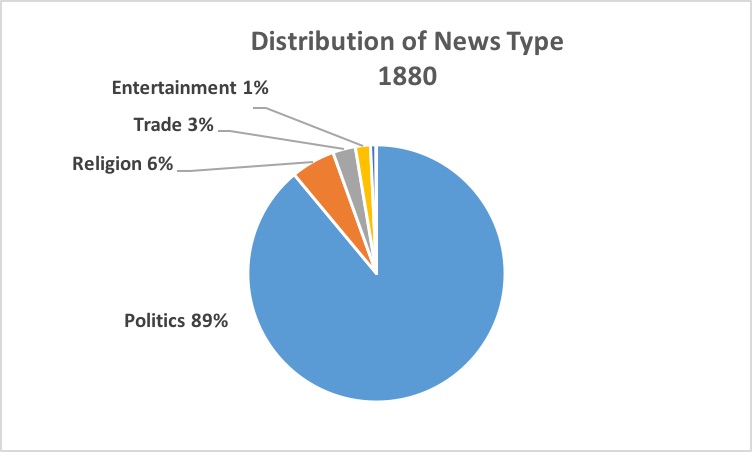}}\hspace{1em}%
  \subcaptionbox{News type by user volume of the top ten visited websites in 2015. \label{fig5:b}}{\includegraphics[width=1.7in]{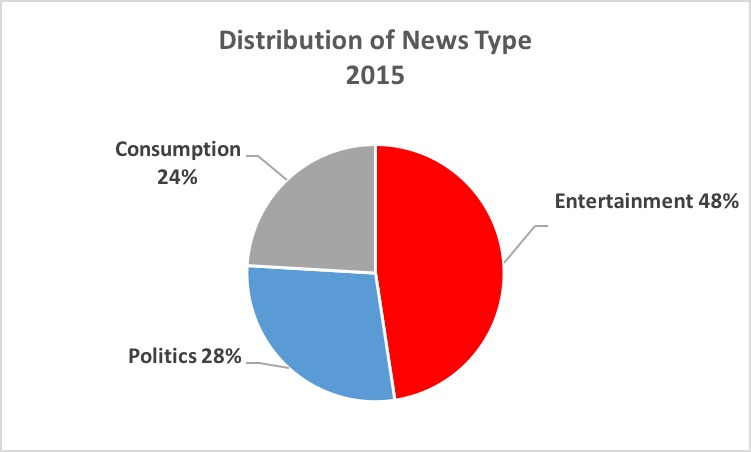}}\hspace{1em}%
\end{figure}
\begin{figure}[H]
 \centering
  \subcaptionbox{\label{fig6:a}}{\includegraphics[width=1.8in]{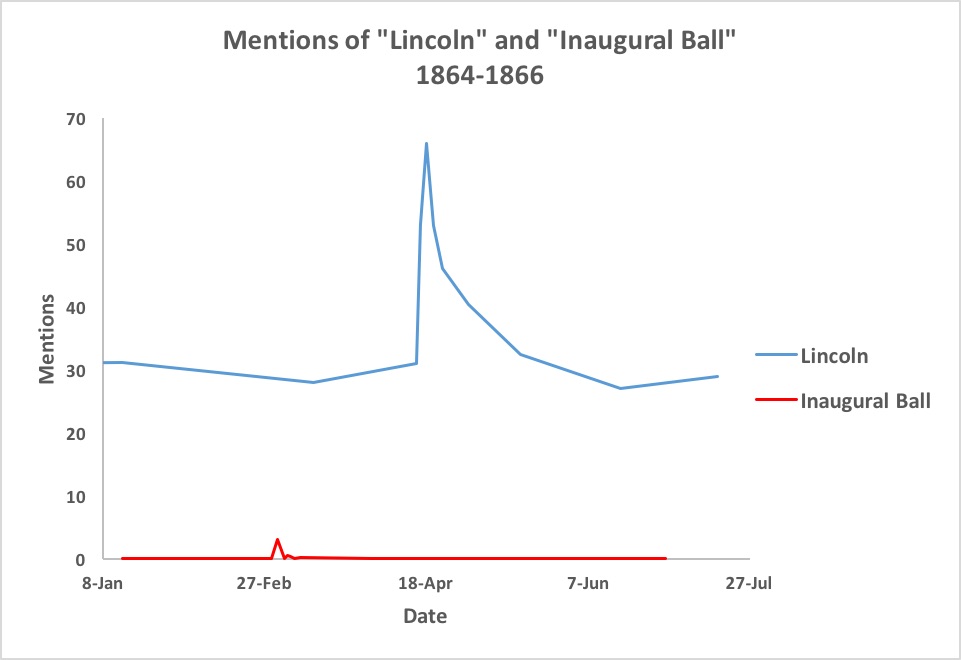}}\hspace{1em}%
  \subcaptionbox{\label{fig6:b}}{\includegraphics[width=1.95in]{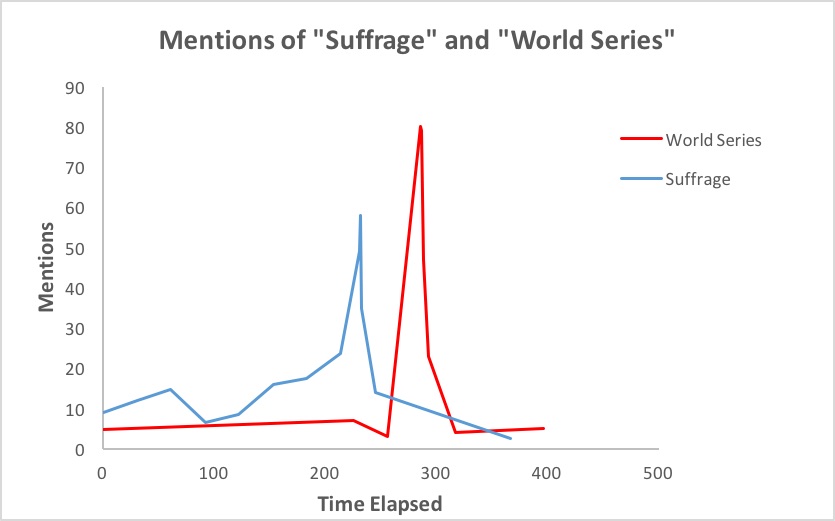}}\hspace{1em}%
  \caption{Comparison of a political news story and entertainment piece of 1865 (a) and 1920 (b).}
\end{figure}

\section{Models}

\subsection{How Information Spreads}
\indent
\indent In order to understand the nature of any piece of information and its spread throughout a network, we first note the ways in which information can travel. Hence forth, we will consider that communication is either \textit{active} or \textit{passive}. Active communication requires focus on the part of both the sharer and receiver of information. A notable example would be newspapers, the journalists and editors of a newspaper must certainly pay attention as they place stories in the next day\textquotesingle s run, and those that consume newspapers have to focus on consuming the text presented. Conversely, some mediums are passive in nature. Examples such as radio and at times television, only require the focus of one of the participants, the one sharing the information. The person receiving may instead focus on such things as, say, driving a car or tasks around the home. A note here should be made about the internet. The internet is distinctly upload-oriented which has often been used as one of its major selling points; the freedom of people to express themselves universally. In addition, the internet is at times both passive and active. There are articles one can read and interactive games, but at the same time streaming services such as Netflix, Spotify, and podcasts. While the internet has not yet branched out largely to those areas traditionally held by passive media, steps are being taken, such as the listening to podcasts in the car rather than radio.\\

\subsection{Previous models and limitations}
\indent
\indent The comparison to diffusion is largely seen in modeling information spread. The model has a basic form of $$\frac{dA(t)}{dt} = \mathit{i}(t)[P-A(t)]$$ In this sense, A(t) are the individuals that have received the information, P is the total population, and \textit{i}(t) is a diffusion coefficient. The diffusion coefficient can be expanded to include an addition factor independent of the number of current individuals pertanent to the information, a multiplicative factor accounting for internal effect of imitation on the spread of the knowledge, or a combination of the two. These models are outlined below.

\[  \mathit{i}(t) = \left\{
\begin{array}{ll}
     \alpha\\
     \beta A(t)\\
     \alpha + \beta A(t)\\
\end{array} 
\right. \]

\indent These coefficients can be implemented in the above equation and can be solved to determine the volume of people prevalent to information at given times. \\

\indent This model effectively maps the increased spread of diffusion given internal and external influence qualities. However, it lacks in versatility and does not account for other factors which will enhance or inhibit the spread of information. \\

\indent In modeling the spread of information in a graph, previous work can most often be categorized by two distinct model types. The Threshold model is defined by a system wherein people begin to adopt a position if a certain ratio of their neighbors adopt it.  Newer solutions falling under the Cascade model suggest an alternative explanation for how information spreads in social networks. In this model type, each node infected with information has a certain probability to infect each of its neighbors. In this manner, one node can actually begin an information cascade, simply by spreading something that is “infectious”, which is analogous to viral media.\\ 
\begin{figure}[H]
 \centering
  \subcaptionbox{\label{fig7:a}}{\includegraphics[width=1.9in]{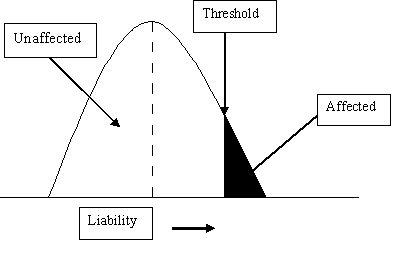}}
  \subcaptionbox{\label{fig7:b}}{\includegraphics[width=1.8in]{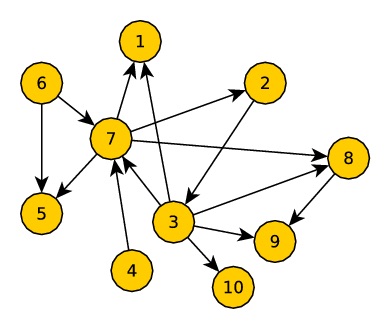}}\hspace{1em}%
  \caption{Basic representation of (a) threshold and (b) cascade models.}
\end{figure}

\subsection{Diffusion Model with Multiple Sources}
\indent
\indent We herein introduce an enhanced model of information diffusion that accounts for the factors lost in the previous model. In order for our model to accurately describe the adoption of a story into mainstream news and track the resulting spread, we take into consideration the following factors:
\begin{itemize}
\item The topical distribution of newsworthy items has changed over time and is not, nor has ever been, equal among topics. In a community oriented towards one subject, an item of that subject will spread quicker than one from a different subject. The topic of the item will affect the spread among a population and is an internal factor.
\item Different items have varying levels of importance to a community as a whole. Some items with high importance, like the assassination of a President, will span the gaps of interest distribution in a population. The importance of a item will affect the spatial spread of the information as well as the speed with which it accomplishes this spread and is an external factor. The importance of an item is determined by a number of characteristics:
\begin{itemize}
\item Novelty or Shock (ability to spark interest without back story)
\item Accessibility (ability of any individual to understand)
\item Attractiveness (impact of news on individual)
\item Compatibility (alignment with societal values and/or acceptable notions)
\end{itemize}
\end{itemize}
These considerations combine to give a total \textit{Relevance Factor} which we will model scaled from 1 to 10.  This relevance factor will determine the rate of spread of the information to the connected nodes of the network. An item with a high relevance factor will have high probability of penetration, velocity, and retention. We remind the reader that an item run through our model has already been categorized as ``news" and therefore exceeds the respective thresholds. Thus, we herein assume an item will propagate to some extent and model the subsequent spread of the item with respect to its relevance factor. \\

\indent Let y(t) denote the number of individuals who are pertanent to a piece of information. In order to accurately model the topical distribution for a given time, we designate a certain percentage of individuals to be ``interested" in each topic based on compiled data of the time. For example, for a model of 1880, 89\% of the individuals will be politically oriented and 7\% will be entertainment. We are not insinuating that real individuals have single-minded interests, but simply that the percentages of interest of a community will change given the relative distribution of the time, modeled by designating percentages of nodes, or individuals. This will in turn account for the respective effect of item topic on the spread.  If a news piece is topically relevant to one community, it will naturally spread between individuals within that community, but also has the potential to leak to individuals of other communities. Essentially, an entertainment individual will still hear and share news of a high profile political story, and vice versa. The key in this cross-community sharing lies in the relevance of the information itself. A low relevance political story will spread slowly in an entertainment network, and a high relevance political story will still spread quickly in a consumption network despite the conflicting interest. This concept can be modeled in the following way. $$\Delta y = R(t)\lbrace\left[\beta_1 y_1(t)+\eta_{12}y_2(t)\right](N_1-y_1(t))+\left[\beta_2 y_2(t)+\eta_{21}y_1(t)\right](N_2-y_2(t))\rbrace \Delta t$$ Or generally, $$\Delta y = R(t) \lbrace \sum_{i=1}^{n}\sum_{\substack{j=1\\j\neq i}}^{n} \left[\beta_i y_i(t)+\eta_{ij} y_j(t)\right](N_i-y_i(t))\rbrace \Delta t$$
\indent Each \textit{i} represents the components of a different topical community. For example, $N_1$ denotes the community interested in entertainment, $N_2$ in politics, etc.. $\beta_1$  represents the rate at which an individual in community 1 will spread an information piece relevant to his community. $\eta_{12}$  represents the rate at which he will spread the information to someone who does not know the information in a different community. $\beta_2$  and $\eta_{21}$ are equivalent variables for an individual in community 2. For a piece of information relevant to one community, it will spread quickest between nodes of this community, slower between those nodes and conflicting nodes, and slowest through two nodes of conflicting interest to the story. Generally, when information \textit{i} is released $\beta_j<\eta_{ji}<\eta_{ij}<\beta_i$ where i and j range from 1 to the number of categories and $i \neq j$. \\

\indent In determining the final number of individuals prevalent to the information in a given time, the function is multiplied by an external relevance factor, R(t). This accounts for information that spans the gaps in community interest. Although as a community we value some news pieces more than others, there are certain viral news stories that bridge these differences. As a people, we are affected by major political changes, influenced by mass entertainment, and impacted by new innovations. Although we are trending away from political news towards technological and entertainment oriented news as a constant influx of information, spikes of important stories will make their way to the individuals of all communities. \\

\indent R(t) can be dynamic in time and allows for consideration of the characteristics described above. In general, items rise and decay exponentially at a speed proportional to their relevance. For example, an item that is novel but not necessarily impactful will have a quick ascent in volume followed by a quick decay. Whereas an item that is novel, impactful, and accessible will have a quick ascent and slow decay corresponding to high penetration and retention. \\

\indent This property is reinforced in volume data collected from real event spread.
	
\begin{figure}[H]
 \centering
  \subcaptionbox{\label{fig8:a}}{\includegraphics[width=1.75in]{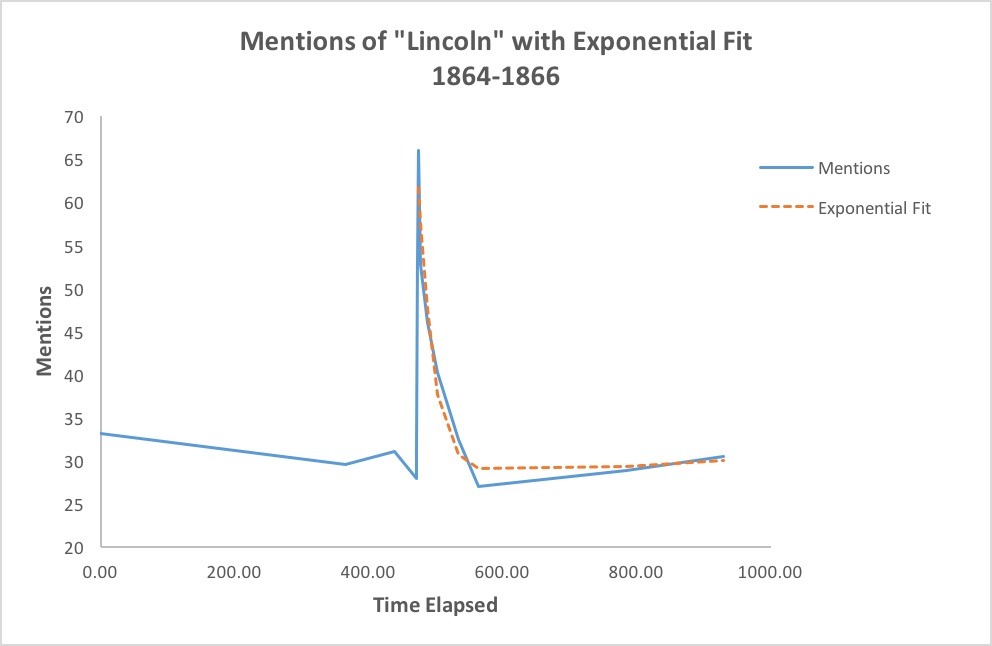}}\hspace{1em}%
  \subcaptionbox{\label{fig8:b}}{\includegraphics[width=1.8in]{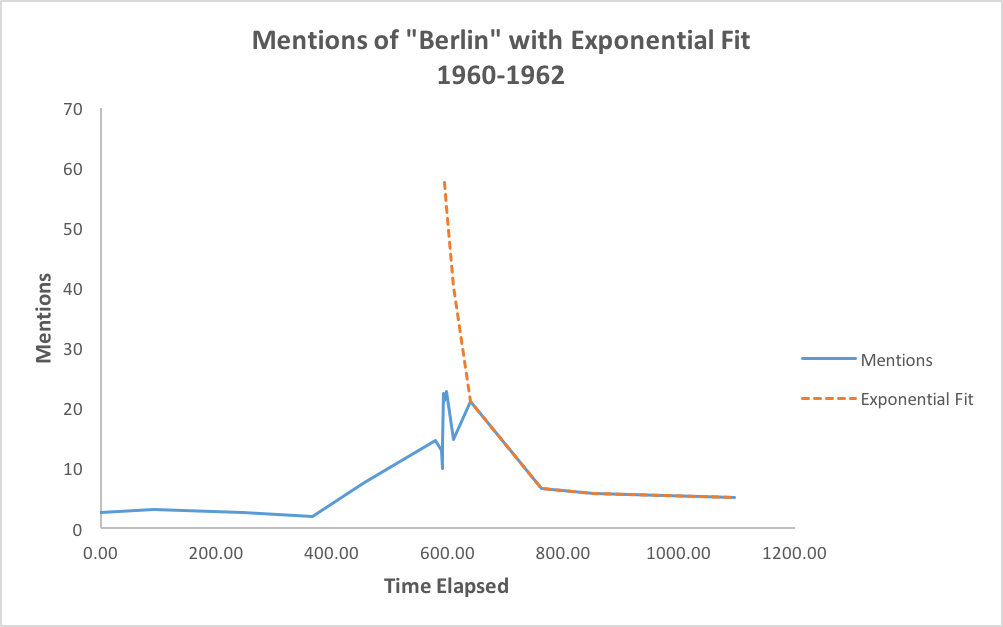}}\hspace{1em}%
  \caption{An exponential function is fit to the decay of mentions over time.}
\end{figure}
\begin{figure}[H]
\centering
\includegraphics[scale=.6]{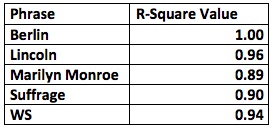} 
\caption{\small The R-Squared values of fits for the aforementioned news stories. An exponential decay function in our model is clearly physically reasonable.} 
\end{figure}
\indent The accuracy of the fit is shown in the above table. Thus far we have shown that the spread of a piece of information is dependent on various internal and external influence factors contributing to a combined relevance factor that varies with time. We have also shown the ability to model and predict the nature of R(t) given the characteristics of the incident information.\\

\subsection{Network Model}
\indent To further our model, we base it in graph theory, with the goal being to observe the flow of information in a population of people connected by different means of communication, or different “types” of edges. In considering the nature of avenues of news, a few considerations are readily apparent. Some forms of communication are distinctly consumption-oriented in nature where one merely receives information from these mediums, such as newspapers, radio, and television. This is represented on a digraph as a directed edge. Conversely, forms of communication that are upload-oriented in nature include telegraphs, the internet, and talking have an opposite that represents itself as merely an edge.\\

\indent Take a node, 0, at time t, (going into an internal state, \(\mathcal{I}_{0,t}\)) with inward connections to nodes $\{$ 1, 2, ..., \(j\) $\}$ and outward state (the state the node is transmitting to its neighbors) represented at time t by \(n_{i,t}\). We represent the absorption of information by:
\begin{align*}
&(\mathcal{I}_{0,t,k} = n_{k,t-1}\cdot P_{N_0})\\
&(\mathcal{T}_{0,t} = \prod _{k=1}^j \left(1 - \mathcal{I}_{0,t,k}\right))\\
&(\mathcal{I}_{0,t} = -\mathcal{T}_{0,t}^2 + 1)
\end{align*}

\indent Given that any \(n_{i,t}\) has a value of 0 or 1, this calculation multiplies 1 or 0 by another 0 or 1, determined by \(P_{N_0}\), a function that semi-randomly decides whether to share something or not. Then, in the next step, we subtract this number from 1, meaning that if 0 notices the information shared by a specific node, it zeroes out the entire equation, if not, it stays at 1. From then, we modify our final result so 1 switched to 0 and vice versa.\\
\indent Now, while this is a method to update the internal “belief” of a node, it does not necessarily signal to other nearby nodes this information. We update their “external beliefs”, what they share to the public, after a gestation period, G, from time \(t_0\), in the following manner, where \(P_{S_0}\) denotes a 0 or 1 in the same sense as \(P_{N_0}\), but for sharing rather than noticing:
$$n_{0,t} = \delta _{t,t_0+G}\cdot P_{S_0}\cdot \mathcal{I}_{0,t}$$
\indent Now, one may note that this is only ever 1 at a single time slice, as it uses the Kronecker delta to denote the time when this is put out into the open, as it is a pulse of information, not a sustained signal. In reference to previous literature, this very much resembles a modified Independent Cascade model for social networks or an SIR epidemic model but with increased intricacy.

\subsection{Network Versatility}

\indent While our model as it stands is already quite versatile and can represent a number of scenarios, a few important problems stand out immediately. While there is a ``gestation'' period for vertices, the model lacks any means of transmitting data from two connected points over a timescale longer than 1 step. To that end, we bifurcate our vertices into two subcategories, “real nodes” and “ghost nodes”. A real node is simply a node described above, it represents a person, and is where information may or may not reside. A ghost node, on the other hand, is mathematically equivalent to a real node, except that \(P_{S_0} = P_{N_0} = 1\). In addition to this specialized nature of it noticing and transmitting data, a ghost node between (0, $n$) also only has two directed connections, one inward from 0, and one outward towards $n$. To expand even further on this model, if the speed of a connection between two nodes is v, the time it takes to transfer between them as v steps (equivalent to things such as newspapers moving on trains), this ghost node would have a gestation period of $v-1$. Finally, to account for the real world scenario of an individual forgetting a piece of news, the internal state of node 0, that has ``believed'' something at time $t_0$, at times after $t_0$, is represented by:
$$(\mathcal{I}_{0,t} = \mathcal{I}_{0,t_0}\cdot P_{V_{0,t}})$$
\indent Note here that \(P_{V_{0,t}}\) is 0 or 1, based on a weighted number generator, however, it changes over time. As \(t-t_0\) increases,  \(P_{V_{0,t}}\) will tend towards 1, as people become more forgetful of information over time. Now, instead of the model simply reflecting people retain information, it has people receiving information, forgetting it, then receiving it again from another source. Much like how people hear news from an initial source, like the newspaper or Facebook, go about their day and forget it, then hear about it again from someone in the home.\\

\section{Simulations}
\indent The parameters mentioned above were placed in a simulation with a few additional modifications. As mentioned in Section 3.3, we categorize news as falling into one of three functions: politics, entertainment, and consumption. The nodes follow the proportional topical distribution of each time era. We only allow data to originate from one source and the nodes themselves can only hold one string at a time until another piece of data overrides it. We run the simulations on two networks representing the cultures of the 1880 and 2015. The differences include the topical proportions, and spread layouts. These correspond to the media categories and sources of the times. Finally, the simulation allows for a random generation of information input (order, type, relevance) or a controlled news input. A few simulations are represented below.\\
\begin{figure}[H]
 \centering
  \subcaptionbox{\label{fig10:a}}{\includegraphics[width=3in]{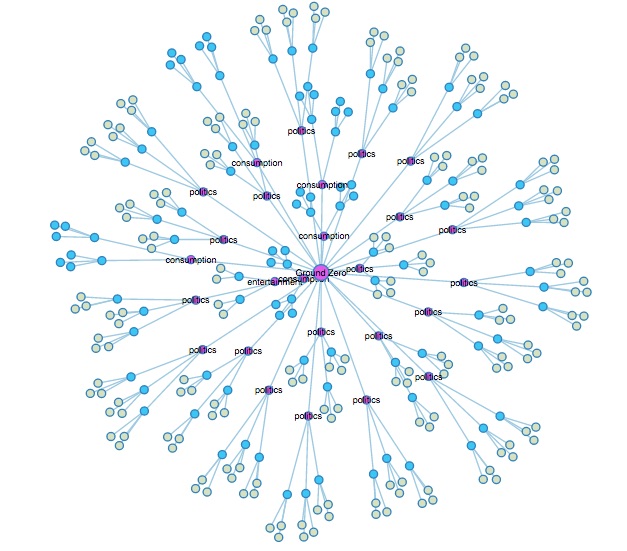}}\hspace{1em}%
  \subcaptionbox{\label{fig10:b}}{\includegraphics[width=3in]{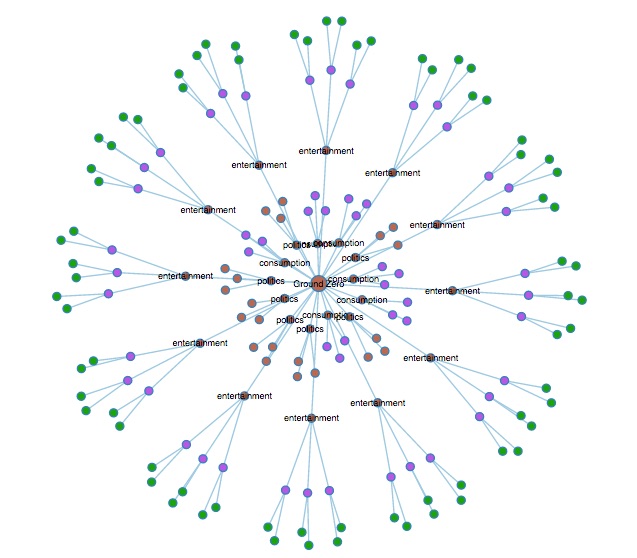}}\hspace{1em}%
  \caption{Randomly generated news simulated in the 1880 (a) and 2015 (b) model. Notice the differences in community proportions and graph connections.}
\end{figure}

\subsection{Validity of Simulation}
\indent
\indent For the validity of the model, we simulate news of varying topic and relevance and compare it with real world data collected. 
\begin{figure}[H]
\centering
\includegraphics[scale=.25]{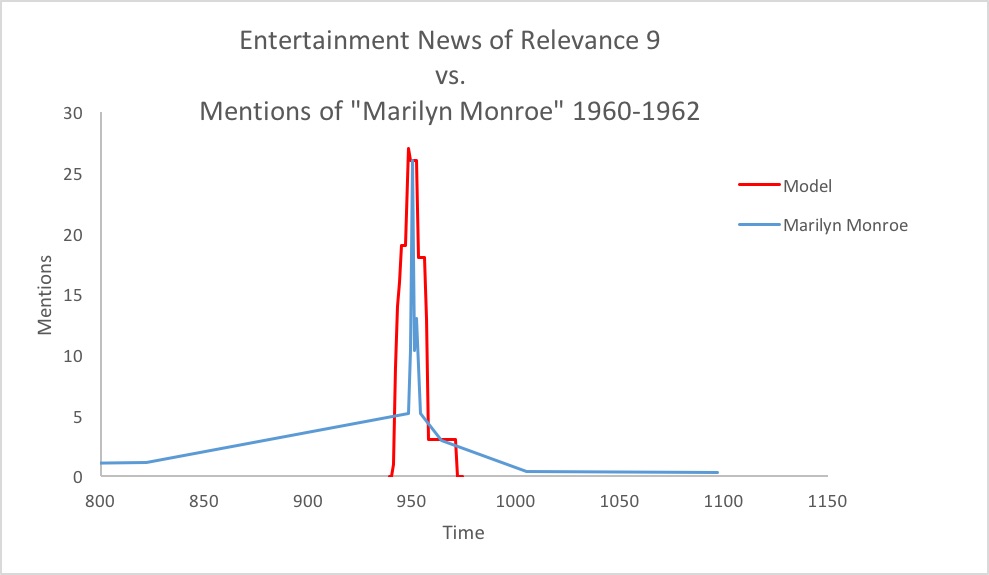} 
\caption{\small The volume of mentions in time of an entertainment item of relevance 9 and the data compiled around Marilyn Monroe's death.} 
\end{figure}
\indent As seen, the components of the real data are accurately reflected in the model. This news was highly relevant with high novelty, attractiveness, and accessibility. The death of Marilyn Monroe would have clearly had a high relevance factor in the time and that is reflected in the model. In a similar sense, our model for current spread and distribution reflects the news of Alan Rickman\textquotesingle s death in January 2016 when compared with volume data from Google Trends. 
\begin{figure}[H]
 \centering
  \subcaptionbox{Model simulation of entertainment news (8) in the 2015 model.\label{fig12:a}}{\includegraphics[width=2.2in]{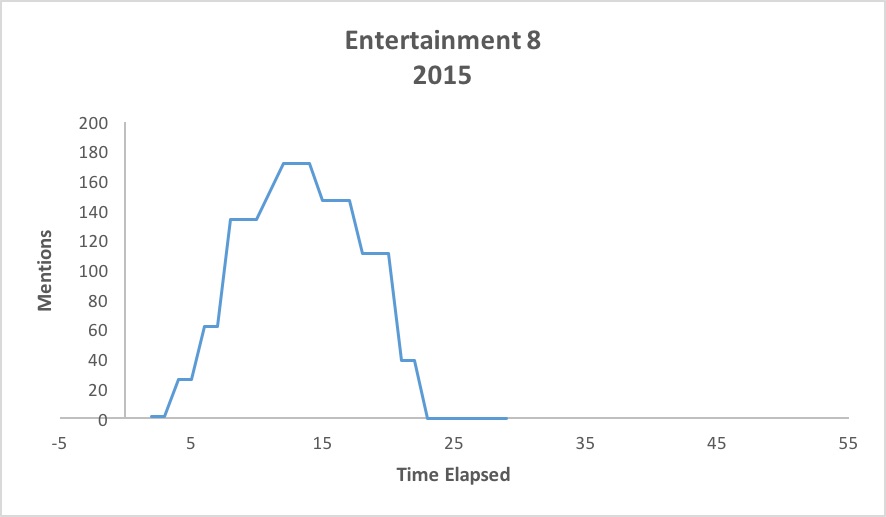}}\hspace{1em}%
  \subcaptionbox{Real world trend of mentions of ``Alan Rickman'' over a given time.\label{fig12:b}}{\includegraphics[width=2.2in]{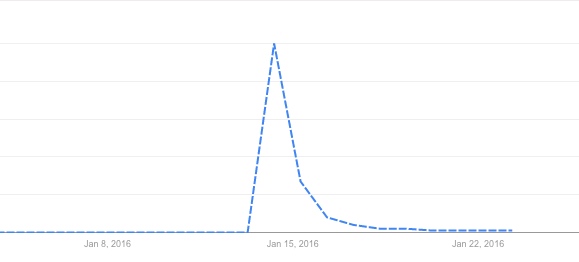}}\hspace{1em}%
\end{figure}
\indent From these comparisons, we validate our model.\\

\subsection{Robust Nature of Simulation}
\indent
\indent To test the robust nature of our model, we evaluate the predicted propagation of current news in past networks and vice versa. Take, for example, the announcement of Osama Bin Laden's capture and death. This would be considered political news of relevance 10. Also, consider the announcement of Kim Kardashian\textquotesingle s pregnancy, an entertainment relevance of 6. We then test this information in our 1880 network to model its spread. The results are shown below.
\begin{figure}[H]
\centering
\includegraphics[scale=.2]{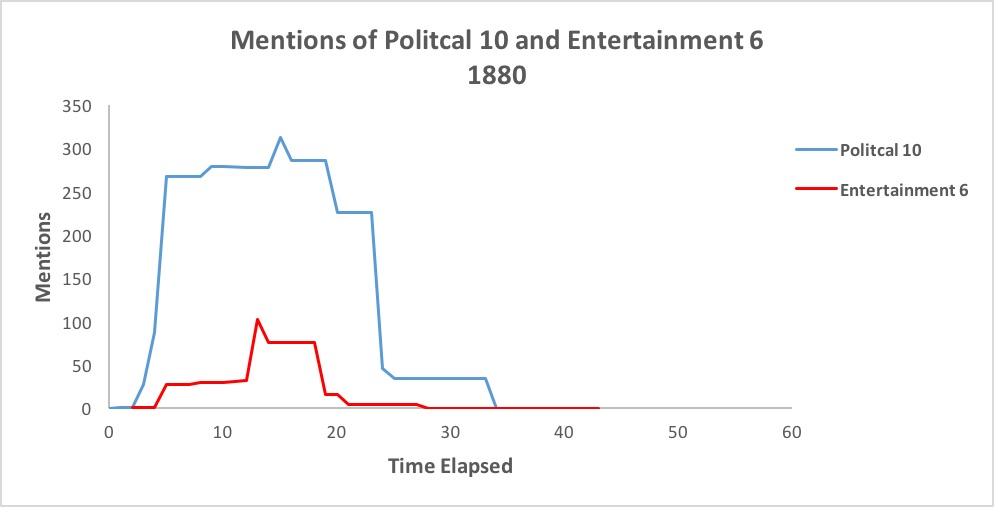} 
\caption{\small Simulation run on 1880 network with news of political relevance 10 and entertainment relevance 6.} 
\end{figure}
\indent To get a sense of the accuracy of the data found, notice the similarities between real volume data of Abraham Lincoln's assassination and news of the inaugural ball. The maximum height of mentions is significantly lower for the entertainment news as well as the retention period. We confidently state that the model well reflects real scenario data across time periods.
\begin{figure}[H]
\centering
\includegraphics[scale=.2]{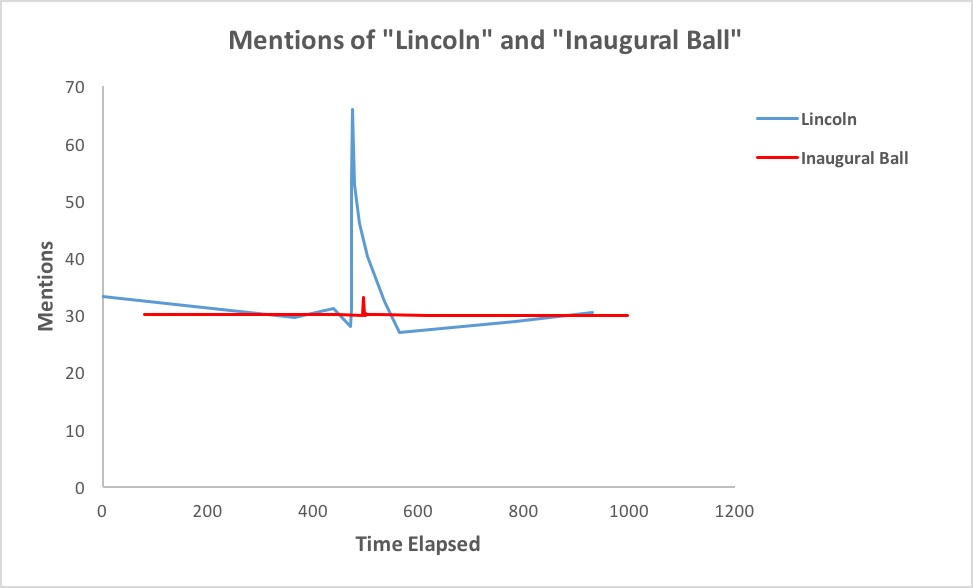} 
\caption{\small The volume spread of Lincoln's assassination and his earlier inaugural ball. The inaugural ball data has been shifted linearly in time to see direct comparison.} 
\end{figure}

\section{The Future of Communication}
\indent
\indent In looking to the future, we first note on the source changes over time. As mentioned, there are active and passive news sources as well as upload and consumption based sources. Upload-oriented networks, in and of themselves, are capable of doing all the things a consumption-oriented network can and more, and will dominate when the resources needed to sustain both are present. However, as those modes of communication fight with each other for relevance, active and passive media act differently, complimenting each other. While there is indeed a significant amount of area in our media consumption space that the two will vie for, active media has a higher volume of communication in a given period of time, while passive media has occupancy space advantages. For example, passive media can be consumed while driving or completing other tasks where active cannot. In this sense, the two complement each other and neither can dominate entirely. As it stands, the internet is immune to competition as it encompasses qualities of all the abovementioned. A new source of media must either improve an existing feature or remove flaws that inhibit information spread of the future.\\

\indent A common occurrence in science fiction is the idea of a neural uplink, transmission of information between minds directly. It is inherently faster than the internet, as it drastically shortens the gestation period. Additionally, it combines all types of communication expertly, effectively layering the models of interaction. The components of face-to-face conversation can be enhanced by the speed of internet connection. A model of this mode would include multi-faceted edges that would behave in compounding fashion. As it trends, we would expect the nodes to further represent entertainment as news, but with this new wave of technology, it is impossible to predict if individuals will lose community sense and begin to function on a personal premise given the immediate availability and personal nature of media associated with a neural uplink. \\

\section{Sensitivity Analysis}
\indent
\indent Our base model has a few fixed parameters in simulating the spread of information throughout a network. We address the impact of these parameters by varying them in the following ways.\\

\subsection{Competing Sources}
\indent 
\indent To study the propagation of news, we limited the number of source nodes to one. In this way we effectively tracked the singular path of this item. We have altered our simulation to include two competing sources of news in an interconnected system. When two sources of information of the same type are released simultaneously into the network, we follow the subsequent spread.\\

\subsubsection{Varying Levels of Importance}
\indent We assume that there is a 50/50 chance for dominance when the two pieces of data meet at a node. Upon setting up this model, two simulations were run, one where the two pieces of information are equally believable, and one where one is much more believable than the other, or having higher relevance. The results are just as one would expect, with the first splitting the graph equally and the second being overrun by the more popular belief. In this sense, we can see how misinformation that suits our preconceived notions spreads, it is more believable to us and thus blots out real information in the network.
\begin{figure}[H]
 \centering
  \subcaptionbox{Equal value. \label{fig15:a}}{\includegraphics[width=1.8in]{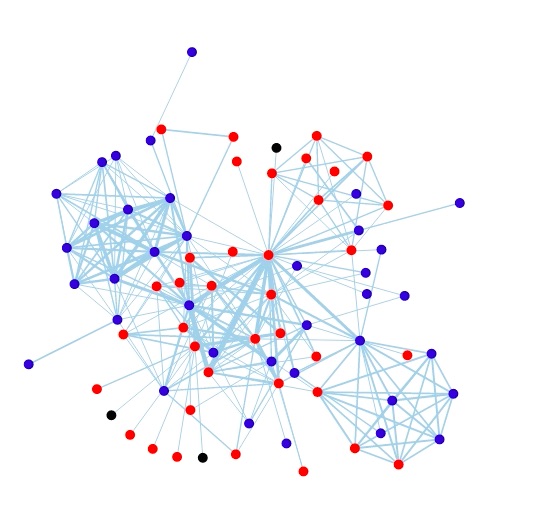}}\hspace{1em}%
  \subcaptionbox{Unequal value.\label{fig15:b}}{\includegraphics[width=1.8in]{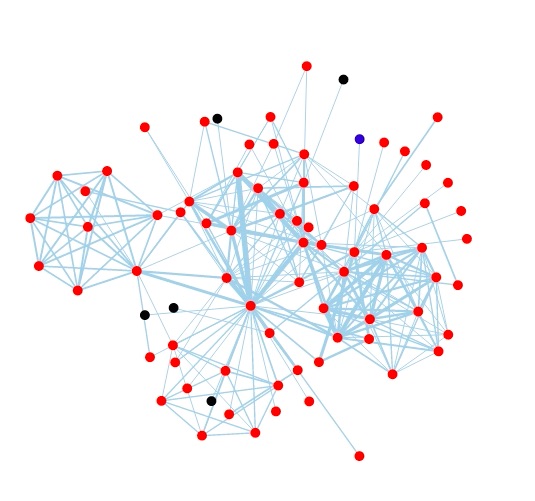}}\hspace{1em}%
  \caption{The spread of information of higher value spreads quicker and further than that of a lower value. This is amplified if the competition is increasingly uneven.}
\end{figure}

\subsubsection{Varying Initial Connectivity}
\noindent For two sources of equal believability, the placement of the initial prevalent nodes hold increasing importance. This is akin to equal news beginning in a rural village as opposed to a large city. We find that the item beginning in a more connected node reaches a larger amount of subsequent nodes in a shorter time. This notion is physically reasonable in that less people will be prevalent to and, thus, willing to accept, town gossip over city news. 
\begin{figure}[H]
 \centering
  \subcaptionbox{Intial sources. \label{fig16:a}}{\includegraphics[width=2.3in]{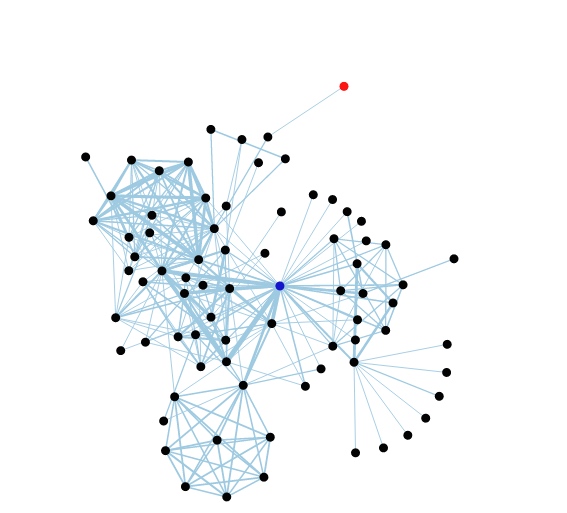}}\hspace{1em}%
  \subcaptionbox{Spread after time.\label{fig16:b}}{\includegraphics[width=1.8in]{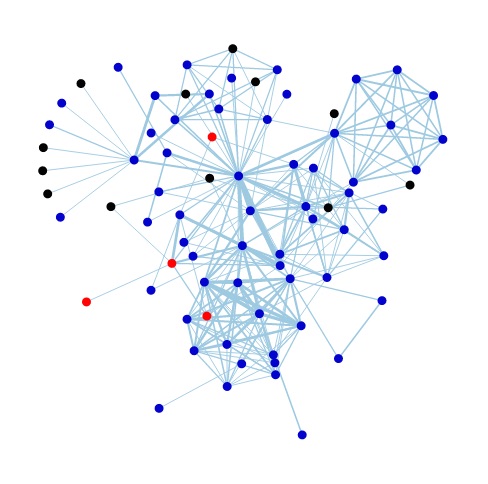}}\hspace{1em}%
  \caption{It is clear to see the respective spreads of red and blue information items and how their initial position or connections influenced this.}
\end{figure}
\begin{figure}[H]
\centering
\includegraphics[scale=.25]{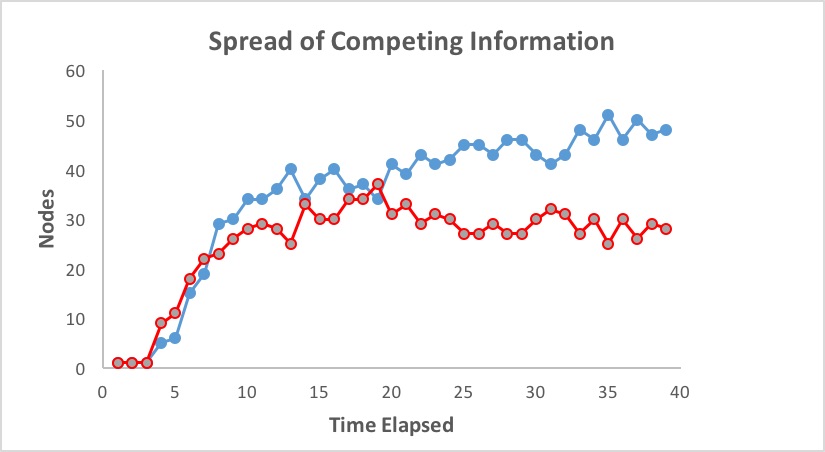} 
\caption{\small Red and Blue information pieces with equal relevance factors of 10 were released simultaneously. Blue was released in a more connected hub.} 
\end{figure}
\
\indent Recall the internet volume data surrounding the news of Osama Bin Laden\textquotesingle s capture and death (Figure 3). The distribution closely resembles the spread of competing information (Figure 17). As ``Osama Bin Laden" became a more relevant phrase, its volume and spread overtook that of a similarly oriented phrase ``Moammar Gaddafi". We further prove the validity of our model through this comparison.\\

\subsubsection{Varying Connection Distances}
\indent
\indent Finally, we alter the model to vary the distance between connections, making it so it takes longer to move across certain edges than others. There are now groupings of connectivity that vary in size and spread which allows simulations of differing topologies in both of our graph and the physical world. For example, an interconnected island shares a network with the mainland. It takes a fair amount of time to move information between them by boat, but the said information propogates much faster both on the “island” and the “mainland” themselves. This differs from the previous simulation in that the connective speeds as well as the connectivity volume are varied.\\
\indent As before we consider two items that are equally relevant or believable. We find that the information originating on the island remains isolated to that island, whereas the mainland information dominates the mainland. Next, we also consider news beginning with a higher relevance or believability, this time on the less connected island, and news with a lower value, now beginning on the mainland. The two views establish an equilibrium with roughly the same adopting nodes. 
\begin{figure}[H]
 \centering
  \subcaptionbox{\label{fig16:a}}{\includegraphics[width=2.1in]{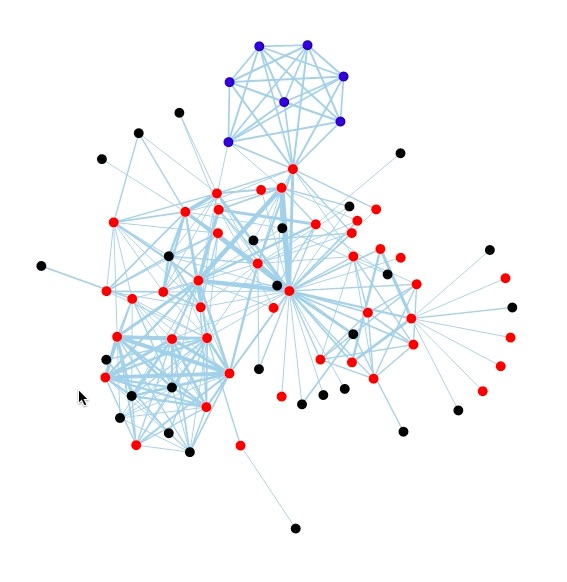}}\hspace{1em}%
  \subcaptionbox{\label{fig16:b}}{\includegraphics[width=2.1in]{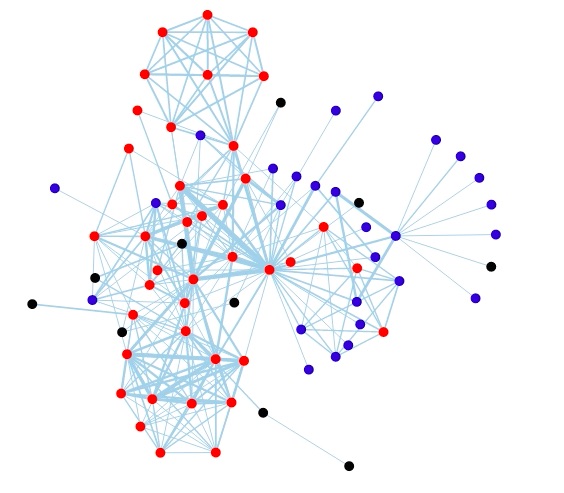}}\hspace{1em}%
  \caption{The resulting spread of two pieces of information beginning in an isolated (blue) and connected (red) community, respectively. (a) represents equal relevance information simultaneously spread and (b) higher relevant information (blue) released in the isolated community.}
\end{figure}
\indent Finally, we note that if the data from the mainland is more relevant, the data on the island dies out quite quickly, and becomes a footnote in history. In these ways we can explicitly determine that physical isolation, much like social, is detrimental to the spread of information. If news starts out isolated by the topology of our social networks, it must make up for it in other ways simply to remain relevant.\\
\begin{figure}[H]
\centering
\includegraphics[scale=.27]{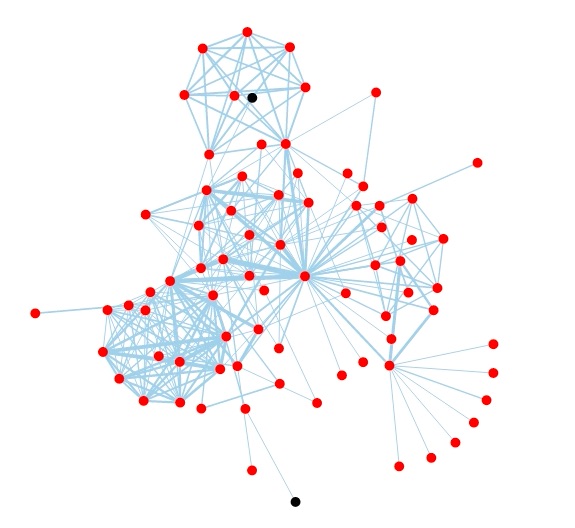} 
\caption{\small Higher relevance information (red) is initiated in a connected community simultaneously with information (blue) initiated in an isolated community.} 
\end{figure}

\subsection{Optimization of Spread}
\indent From these simulations we can also make statements about an optimaztion of spread strategy. This information is useful for individuals or groups interested in conveying information to a maximum number of people. This analysis furthers the expected belief that in order to maximize spread, the information needs to have high relevance, topical importance, and a carefully selected initial source. A node with many edges is not necessarilly more connected if the surrounding nodes are weak. Ideally, a piece of information needs to be novel, accessible, attractive, and compatible. It needs to, in some way, be able to be characterized with the dominant topic of the time. Finally it needs to be initiated in a carefully selected initial place with the highest connectivity--both in personal node connections and global community connections. In these ways, the spread of an information piece an be optimized.

\section{Conclusion}
\indent Throughout the development of our model, one thing is clear: there is a definite relationship between the importance of a piece of information and its resulting spread. However, this importance is a combination of many factors separately intrinsic to the item and the community. We have extensively shown that our model can account for changes in these factors and accurately reflect the penetration, spread, and retention of the information. Although all news items will spike and decay in varying levels, they follow a consistent model. Thus, by examining the spread of an item, we can extrapolate the underlying factors that contribute to its relevance. Conversely, by knowing the characteristics of an information piece, we can predict the resulting spread. Finally, we can use this knowledge to strategically spread an information item to the furthest bounds. \\

\subsection{Strengths and Weaknesses in the Models}
\indent
\indent In total, our model is extremely multifaceted and inclusive. The diffusion model allows for varying levels of importance to be directly seen in the resulting volume of mentions. It is easily translated for an increasing number of interest categories or types of news and models the interactions of a news piece between the categories. The graph model excels in versatility: spatial distance can be represented based on assigning velocities to edges, different mediums can be assigned for the propagation of information based on velocities and direction of edges. For the model as a whole, the immense versatility allows us to see clear relationships between information with different levels of importance based on a number of inherent characteristics and the respective spread of that item.\\

\indent However, our model is not without weakness. We are unable to model true preferences of individuals, only the preferences of the network as a whole. The connections are generated randomly so the more realistic scenario of having more connections with like-minded people within certain communities is lost, even if the overarching preference of the community is preserved. The model relies on discrete time values and some elements of weighted randomness in deciding how information is noticed and shared. In addition, our model involves a static graph in that no nodes appear or disappear and no edges change. \\

\subsection{Supplements and Future Work}
\indent
\indent Although this model and the analysis of communication is far-reaching, further work on this topic may be done to move closer and closer to real world simulation. First and foremost, one could fully account for the changing nature of human interactions over time. It would further the model to include the proportional array of ways an individual communicates and how that has changed over time. Although individuals and their relationships are discrete, synthesizing it with continuous time would improve this model. Finally, a way to accurately vary learning a news item from a mass media source versus a personal connection  in order to map the resulting internal effect on the spread of that item by the individual would serve to further the model.

\newpage
\section{References}
\bibliography{ICM Model 1}

\noindent Baughman, James. "Television Comes to America, 1947-57." Television Comes to America, 1947-57. Illinois State Library, n.d. Web. 31 Jan. 2016.\\

\noindent Beckford, Martin, and Graeme Paton. "Royal Wedding Facts and Figures." The Telegraph. Telegraph Media Group, 29 Apr. 2011. Web. 31 Jan. 2016.
"Chronicling America." News about Chronicling America RSS. Library of Congress, n.d. Web. 31 Jan. 2016.\\

\noindent Bell, Allan. The Language of News Media. Oxford, UK: Blackwell, 1991. 1-23. Print.\\

\noindent Deutschmann, P. J., and W. A. Danielson. "Diffusion of Knowledge of the Major News Story." Journalism and Mass Communication Quarterly 37.3 (1960): 345-55. Web.\\

\noindent Edmonds, Rick. "State of the News Media 2015 — A New Ranking of Digital Sites." Poynter Institute, 29 Apr. 2015. Web. 31 Jan. 2016.\\

\noindent "Feature Radio In The 1930s." PBS. PBS, n.d. Web. 31 Jan. 2016.

\noindent Geroski, P.a. "Models of Technology Diffusion." Research Policy 29.4-5 (2000): 603-25. Web. 29 Jan. 2016.\\

\noindent "Google Trends - Web Search Interest - Worldwide, 2004 - Present." Google Trends. Google, n.d. Web. 31 Jan. 2016.\\

\noindent Granovetter, Mark. "Threshold Models of Collective Behavior." American Journal of Sociology 83.6 (1978): 1420-443. Web. 31 Jan. 2016.\\

\noindent Graph Representing Social Network. N.d. Independent Cascade Model of Information Diffusion. Web. 1 Feb. 2016.\\

\noindent "Internet Users." Number of (2015). Internet Live Stats, n.d. Web. 01 Feb. 2016.\\

\noindent Kempe, David, Jon Kleinberg, and Éva Tardos. "Influential Nodes in a Diffusion Model for Social Networks." Automata, Languages and Programming Lecture Notes in Computer Science (2005): 1127-138. Web. 31 Jan. 2016.\\

\noindent Kempe, David, Jon Kleinberg, and Éva Tardos. "Maximizing the Spread of Influence through a Social Network." Proceedings of the Ninth ACM SIGKDD International Conference on Knowledge Discovery and Data Mining - KDD '03 (2003): n. pag. Web. 31 Jan. 2016.\\

\noindent Lotan, Gilad. "Breaking Bin Laden: Visualizing the Power of a Single Tweet - SocialFlow." SocialFlow Atom. N.p., 06 May 2011. Web. 31 Jan. 2016.\\

\noindent Printed. Statistics on Radio and Television, 1950-1960; Statistical Reports and Studies 23 (1963): n. pag. Print.\\

\noindent Saito, Kazumi, Ryohei Nakano, and Masahiro Kimura. "Prediction of Information Diffusion Probabilities for Independent Cascade Model." Lecture Notes in Computer Science Knowledge-Based Intelligent Information and Engineering Systems (n.d.): 67-75. Web. 31 Jan. 2016.\\

\noindent Schelling, Thomas C. "Dynamic Models of Segregation." The Journal of Mathematical Sociology 1.2 (1971): 143-86. Web. 31 Jan. 2016.\\

\noindent United States. Census Bureau. Statistics of Newspapers and Periodicals in the United States. N.p.: n.p., n.d. Web. 30 Jan. 2016.\\

\noindent The Washington Post n.d.: n. pag. Historical Archive. Web. 30 Jan. 2016.\\

\noindent Threshold Model of Disease. N.d. Hawaii Genetics Program. Web. 01 Feb. 2016.\\

\noindent Yang, Jaewon, and Jure Leskovec. "Modeling Information Diffusion in Implicit Networks." 2010 IEEE International Conference on Data Mining (2010): n. pag. Web. 30 Jan. 2016.\\

\noindent "The Year 1960 From The People History." What Happened in 1962 Inc. Pop Culture, Prices and Events. N.p., Sept. 2004. Web. 31 Jan. 2016.\\

\noindent Zafarani, Reza, Mohammad Ali Abbasi, and Huan Liu. "Information Diffusion in Social Media." Social Media Mining: An Introduction. N.p.: Cambridge UP, 2014. N. pag. Web.\\

\end{document}